\documentclass[openacc]{rsproca_new}
\usepackage{lineno}
\newcommand{\newpart}[1]{{#1}}
\usepackage{comment}

\newcommand{\Pe}{{\rm Pe}}
\usepackage{multicol}
\jname{rsif}
\Journal{J R Soc Interface\ }

\begin{document}

\title{Optimal switching strategies for navigation in stochastic settings}

\author{Francesco Mori$^1$ and L. Mahadevan$^{2,3}$}
\address{$^1$ Rudolf Peierls Centre for Theoretical Physics, University of Oxford, Oxford OX1 3PU, United Kingdom\\ $^2$  John A. Paulson School of Engineering and Applied Sciences, Harvard University, Cambridge, Massachusetts 02138, USA\\ $^3$ Departments of Physics, and Organismic and Evolutionary Biology, Harvard University, Cambridge, Massachusetts 02138, USA}


\subject{behaviour, biophysics}



\corres{L Mahadevan\\
\email{lmahadev@g.harvard.edu}}


\begin{abstract}
When navigating complex environments, animals often combine multiple strategies to mitigate the effects of external disturbances. These modalities often correspond to different sources of information, leading to speed-accuracy trade-offs. Inspired by the intermittent reorientation strategy seen in the behavior of the dung beetle, we consider the problem of the navigation strategy of a correlated random walker moving in two dimensions. We assume that the heading of the walker can be reoriented to the preferred direction by paying a fixed cost as it tries to maximize its total displacement in a fixed direction. Using optimal control theory, we derive analytically and confirm numerically the strategy that maximizes the walker's speed, and show that the average time between reorientations scales inversely with the magnitude of the environmental noise. We then extend our framework to describe execution errors and sensory acquisition noise. As a result, we provide a range of testable predictions and suggest new experimental directions. Our approach may be amenable to other navigation problems involving multiple sensory modalities that require switching between egocentric and geocentric strategies.
\end{abstract}



\maketitle

\section{Introduction}
 Navigation in noisy environments is fundamental for animal survival in many situations \cite{freas2022basis,wiltschko2023animal,menti2023towards}. The inherent stochasticity in such tasks emerges from a combination of limited sensory capabilities, complex environments, and execution errors. Given a finite cognitive capacity, animals have to optimally allocate their resources to simultaneously process signals, elaborate a navigation strategy, and execute motion. The extreme examples of how this set of tasks is brought about are commonly known as \emph{egocentric} and \emph{geocentric} strategies.  A well-known example of the first is seen in the desert ant \emph{Cataglyphis} \cite{wehner1996visual} which relies on continuously updating an internal estimate of the location by integrating sensory and motor information. These \emph{egocentric} strategies work well by allowing for rapid movement in space but are prone to error accumulation over time due to fluctuating environments and limited memory. At the other extreme is the case of a map-equipped human who relies on \emph{geocentric} information, using landmarks cues to adjust their strategy. Since this requires probing the environment in real time, a geocentric scheme is slow but more robust to external noise and execution errors. More often than not, one sees a combination of egocentric and geocentric schemes across species \cite{wehner1996visual,peleg2016optimal,haberkern2016studying}.

\begin{figure}
    \centering
    \includegraphics[width=0.75\columnwidth]{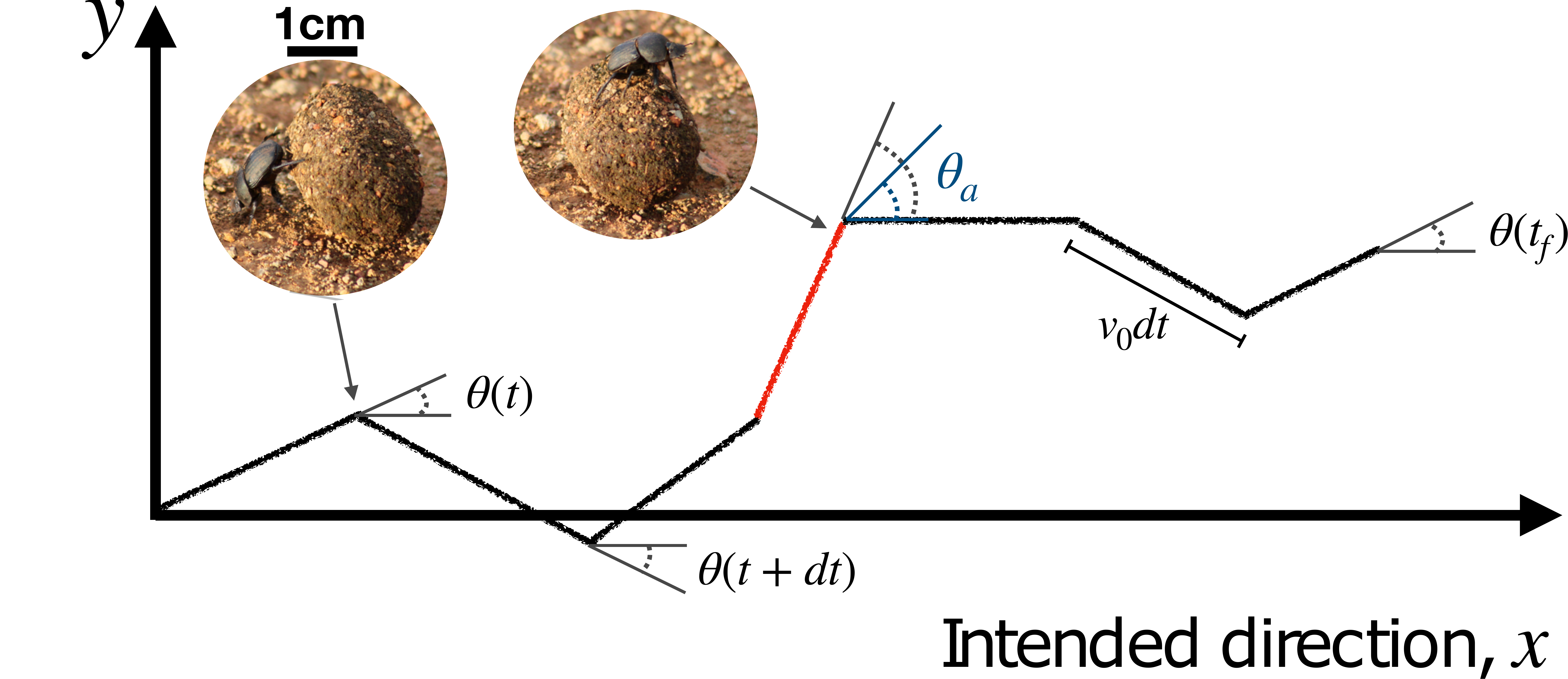}
    \caption{Schematic representation of the navigation model. The agent (e.g. a dung beetle) tries to maximize the displacement in the $x$ direction. It switches between rolling phases, during which it moves at speed $v_0$, and reorientation phases (red line), where the direction $\theta$ is reset to the preferred angle $\theta\!=\!0$. The reorientations are triggered when $|\theta|>\theta_a(z)$, where $z=x_cD/v_0$ is a dimensionless measure of noise (see text for details). Images of the dung beetle in the rolling phase and reorientation phases are from \cite{image}. \label{fig:traj}}
\end{figure}

An example of strategy integration is the motion of dung beetles \cite{byrne2011visual,baird2012dung,foster2021light,dacke2021dung}. In an attempt to escape from the zone of high competition close to a fresh dung pile, dung beetles roll a spheroidal ball of nutrient-rich dung in a fixed direction away from the pile. To persistently move in a straight line, they alternate between rolling and reorientation phases. During the rolling phase, using their dorsal eyes, they acquire directional information from patterns of polarized light and walk backwards, pushing the ball with their hind legs \cite{byrne2011visual}. During the reorientation phases, the beetles climb on top of the ball and dance on it, presumably using long-range cues to correct their heading \cite{video}, see Fig.~\ref{fig:traj}. These reorientation events are thought to be triggered by the accumulation of deviations from the preferred direction due to a combination of sensing and execution errors associated with environmental perturbations and dung ball asphericity.

Experiments on dung beetles indicate that rougher landscapes result in more frequent corrections \cite{baird2012dung,dacke2021dung}, suggesting an underlying optimality principle in decision making. Indeed, a strategy involving frequent course corrections leads to straighter paths, but more time is spent reorienting. Conversely, without external cues, beetles quickly lose their direction \cite{khaldy2019effect}. This type of speed-accuracy trade-off is best analyzed using the mathematical framework of optimal control theory, which allows to determine optimal strategies and their performance.

Motivated by these examples, here we consider the motion of a sentient agent undergoing correlated random walks \cite{codling2008random} that can switch between egocentric and geocentric strategies to maximize its speed in a given direction. The optimal protocols for this model were first investigated numerically in Ref.~\cite{peleg2016optimal}. In this article, we extend these numerical approaches using optimal control theory \cite{bellman1965dynamic,de2023resetting}, to derive an analytic framework for different strategies in the presence of different sources of noise as well as different models of agent-environment interaction. We derive exact formulae for the maximal speed of an agent as a function of the noise magnitude and investigate analytically the effect of execution and measurement errors.

\section{Mathematical model}

We consider an agent moving in a plane with a position $(x(t),y(t))$ with constant speed $v_0$, and heading direction $\theta(t)\!\in\![-\pi/2,\pi/2]$. The position and orientation of the agent then evolve according to 
\begin{equation}
    \dot{x}(t)=v_0\cos[\theta(t)]\,,
    \dot{y}(t)=v_0\sin[\theta(t)]\,,\dot{\theta}(t)\!=\!\eta(t) \quad\label{eq:model_0}
\end{equation} 
corresponding to an active Brownian particle \cite{basu2018active},  where $\eta(t)$ is zero-mean Gaussian white noise with $\langle\eta(t)\eta(t') \rangle\!=\! 2D\delta(t-t')$ and $D\!>\!0$ is the rotational diffusion \footnote{A more appropriate description of the noise would be via a von-Mises distribution that correctly accounts for the periodicity of $\theta$, but for low noise levels, as assumed here, we can neglect the difference between the two}. This type of correlated random walk has been widely employed to describe animal navigation towards a target \cite{cheung2007animal,bailey2018navigational,yin2020simulating,bijma2021sisyphus,yin2022mathematical}, and  assumes that translational Brownian motion is small relative to rotational Brownian motion. This approach has also been used to describe the active motion of Janus particles \cite{walther2008janus}, a type of artificial microswimmer, which when subject to an external field can then be used to control its orientation and position. In all cases, there is a characteristic time that scales with $1/D$, beyond which the agent will lose its orientational persistence. In order to maximize its total displacement in the $x$-direction, the agent must reset its heading to the preferred angle $\theta=0$. We assume that the reorientation process is fast and can be described in terms of an effective displacement lost by stopping for a time $t_s$ to correct the direction with a fixed cost $x_c\approx v_0 t_s$ for each reorientation.  \newpart{Models of this type that combines diffusive processes with switches between distinct modalities are also commonly used to model search behaviors in foraging within theoretical ecology \cite{viswanathan2011physics}.}

\begin{table}[h!]
    \centering
    \begin{tabular}{ll}
        \hline
        \textbf{Symbol} & \textbf{Description} \\
        \hline 
        $\theta(t)$ & Heading angle of the agent at time $t$. \\
        $x(t), y(t)$ & Position coordinates of the agent in the plane. \\
        $v_0$ & Constant speed of the agent during motion. \\
        $D$ & Rotational diffusion coefficient. \\
        $x_c$ & Effective cost associated with reorientation events. \\
        $\theta_a$ & Threshold angle triggering reorientation. \\
        $z$ & Dimensionless parameter $z = x_c D / v_0$, characterizing noise. \\
        $J(\theta, t)$ & Optimal cost-to-go function. \\
        $D_1$ & Diffusion constant for the observable angular variable $\theta_1$. \\
        $D_2$ & Diffusion constant for the hidden angular variable $\theta_2$. \\
        $D_n$ & Diffusion constant for measurement noise $\theta_n$. \\
        $\ell$ & Observability ratio $\ell = D_1 / D_2$ between two angular variables. \\
        $r$ & Signal-to-noise ratio $r = D / D_n$. \\
                $a$ & Magnitude of the correcting drift during reorientation. \\
        $\text{Pe}$ & Péclet number $\text{Pe} = a / D$, quantifying drift-to-diffusion ratio. \\
        \hline
    \end{tabular}
    \caption{\newpart{Summary of notation used in the paper.}}
    \label{tab:notation}
\end{table}

\newpart{Given the dynamical equations Eq. (2.1), and assuming an initial condition $(x(t_0),y(t_0),\theta(t_0))=(x_0,y_0,\theta_0)$, we can then define a cost function for the time interval $(t_0,t_f)$ as the sum of the (negative) distance traveled along the $x-$axis and the distance lost due to $(N(t_f)-N(t_0))$ reorientation events in that time as
\begin{equation}
    \mathcal{F}_{\theta_0,t_0}[s]=\left\langle -[x(t_f)-x_0] +x_c\left[N(t_f)-N(t_0)\right]\right\rangle\,,
    \label{eq:cost}
\end{equation}
where $N(t)$ indicates the total number of correction events in the time interval $[0,t]$ and $s(\theta,t)$ with $-\pi\leq\theta\leq\pi$ and $t_0<t<t_f$ indicates a reorientation strategy, with $s(\theta,t)=1$ if the agent performs a correction at time $t$ and direction $\theta$ and $s(\theta,t)=0$ otherwise. Here $\langle\cdot\rangle$ indicates the average over all stochastic trajectories $\{\theta(t)\}_{t_0<t<t_f}$, with fixed initial condition $\theta(t_0)=\theta_0$.} A strategy with no reorientation events would lead to the second term vanishing, but the agent would only persist in the initial direction over a distance $\sim v_0/D$ \cite{basu2018active}; conversely frequent reorientations would result in a large  cost, and little progression from the initial state. Thus an optimal strategy is one that minimizes the cost Eq.~\eqref{eq:cost} and depends on the ratio of the resetting cost $x_c$ to the persistence length $v_0/D$. Defining $z\!\equiv\! x_cD/v_0$, we expect that for small (large) $z$, the minimization of first (second) term in Eq.~\eqref{eq:cost} determines the strategy. 


\begin{figure}
    \centering
    \includegraphics[scale=0.6]{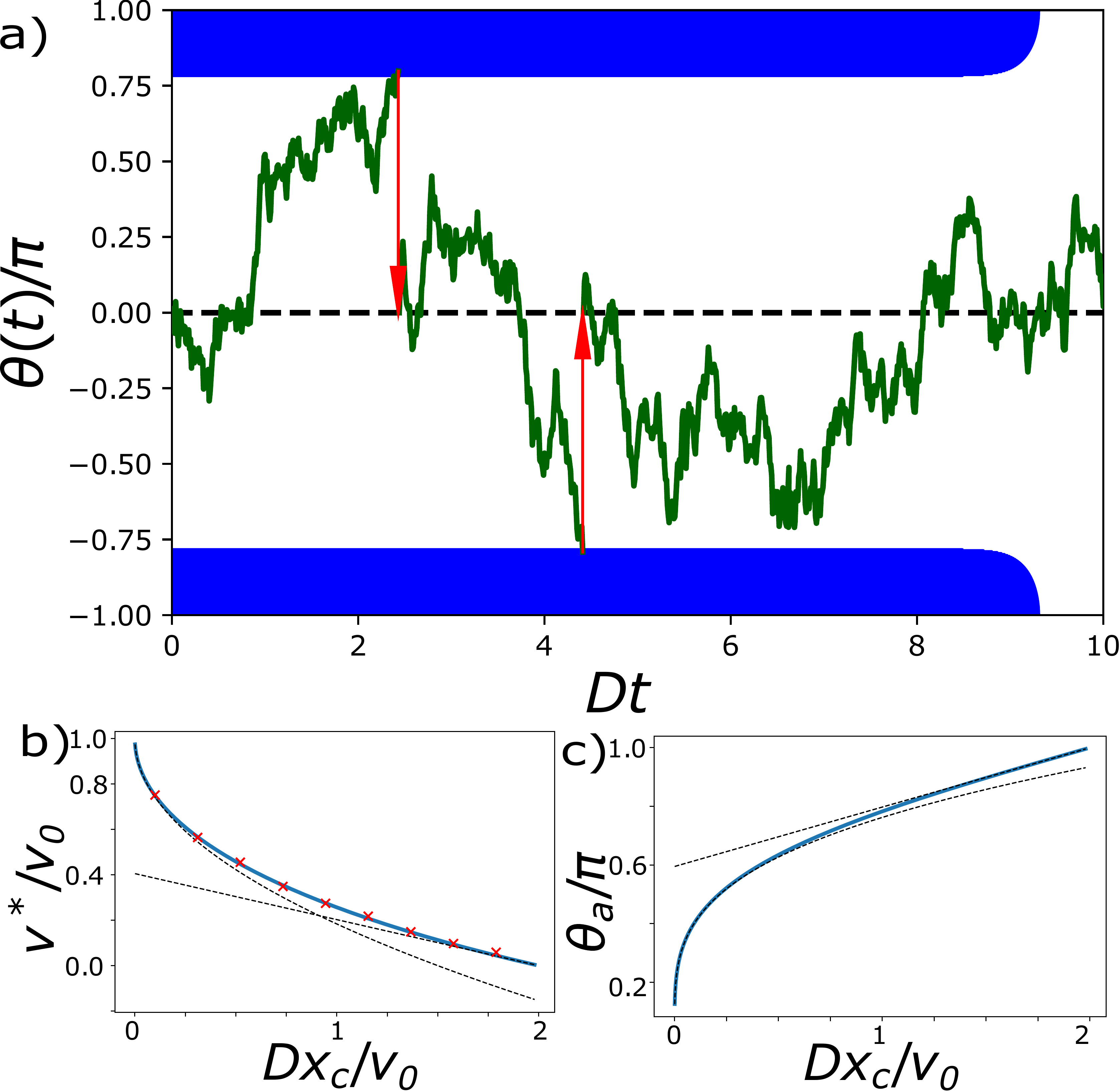}
    \caption{{\bf a)} Optimal restarting policy for $z\!=\!1$ and $Dt_f\!=\!10$, obtained by solving numerically Eq.~\ref{eq:belmann}. The optimal strategy is to reset as soon as the angle $\theta(t)$ touches the blue region, while the system evolves freely in the white region (denoted $\Omega(t)$ in the text). The green line shows a typical trajectory with $N\!=\!2$ resetting events (red arrows). {\bf b)} and {\bf c)} Optimal speed $v^*/v_0$ and optimal activation angle $\theta_a/\pi$ as a function of $z\!=\!Dx_c/v_0$. The blue lines correspond to the numerical solution of Eqs.~\ref{eq:thetaa} and \ref{eq:vstar}, while the dashed lines indicate the asymptotic behaviors discussed in the text. The red symbols in {\bf b)} are the results of Langevin numerical simulations of the optimally controlled dynamics in Eq.~\eqref{eq:model_0}. The average is performed over a single trajectory with $Dt_f=10^4$. \label{fig:omegat}}
\end{figure}

For intermediate values of $z$, a natural approach is to correct the direction when the angular deviation $\theta$ from the target surpasses a threshold $\theta_a$. As we demonstrate below, this turns out to be the optimal strategy when the remaining scaled time $D(t_f-t) \gg 1$. However, it is unclear a priory how to determine the optimal value of $\theta_a$ and the corresponding maximal speed $v^*$. The value of $\theta_a$ should reflect the noise level in the system, though it is unclear whether higher noise should lead to more or less strict control measures, corresponding to smaller or larger activation angles $\theta_a$. Our control theoretic strategy reveals that in noisier environments less stringent control strategies are optimal. This principle also applies when navigation accuracy is compromised by imperfect observations, whether from noisy or incomplete measurements (see Fig.~\ref{fig:v_star_noise} and \cite{supmat}).

\section{Analysis of the model and results}


\begin{figure}
    \centering
    \includegraphics[width=\linewidth]{images_main_text_png/Figure3.png}
    \caption{\newpart{Numerical simulations of optimally-controlled trajectories and their statistical properties for different environmental noise levels $D = 0.1, 0.5, 1.5 $. 
        The trajectories (top row) depict the two-dimensional motion of particles in the $ (x,y) $-plane, with red dots marking reorientation events triggered by exceeding the optimal threshold angle. 
        The boxplots (bottom row) illustrate the distribution of the effective $x$-displacement (left) and lateral displacement, i.e. $|y|$ (right), for each value of $D$. 
        Simulations parameters are $v_0 =x_c = 1$ and $t_f=5$. For the trajectories, we used $n=10$ independent simulations, while for the boxplots we used $n=1000$ to have a better statistics. As expected, the effective displacement in the $x$-direction, which includes both the physical displacement and a cost $x_c$ per reorientation, decreases with increasing $D$. The lateral displacement doesn't change significantly.
    }}
    \label{fig:simulations_trajectories}
\end{figure}

\subsection{Optimal control framework}
Recent studies in non-equilibrium statistical mechanics have shown that resetting a system to its initial position at random times can dramatically improve its search properties \cite{benichou2011intermittent,evans2011diffusion,evans2011diffusionoptimal,evans2020stochastic,kumar2020active,abdoli2021stochastic,sar2023resetting}. Here, instead of following stochastic resetting protocols, we use an optimal-control framework  \cite{de2023resetting} and derive the optimal resetting policy that minimizes the cost in Eq.~\eqref{eq:cost}. Defining the optimal cost-to-go $J(\theta,t)$ as the remaining cost of an optimally controlled system starting at $\theta(t)$, i.e. the cost from time $t$ to the final time $t_f$, \newpart{we may write the optimal value of the cost function 
\begin{eqnarray}
    J(\theta,t)=\min_s \mathcal{F}_{\theta,t}[s]\,,
\end{eqnarray}
where $s$ indicates a reorientation strategy (see \cite{supmat} for details).} \newpart{Using ideas from dynamic programming (see \cite{supmat} for details) we derive an evolution equation for $J(\theta,t)$, which reads}
\begin{equation}
    -\partial_t J(\theta,t)=D \partial^2_{\theta} J(\theta,t)-v_0\cos(\theta)\,, \quad \theta\in \Omega(t)\,, \label{eq:belmann}
\end{equation}
where $\Omega(t)=\{\theta:J(\theta,t)\leq J(0,t)+x_c\}$. Eq.~\eqref{eq:belmann} is a generalization of the Bellman equation \cite{bellman1965dynamic}. We note that the nonlinear gradient term of the standard Bellman equation is absent here because the system is controlled through restarts rather than an external force. Eq.~\eqref{eq:belmann} must be solved backward in time, with final condition $J(\theta,t_f)=0$ and Neumann boundary conditions $\partial_{\theta}J(\theta,t)\!=\!0$ for $\theta\!\in\! \partial\Omega(t)$ \newpart{\cite{de2023resetting}}. The domain $\Omega(t)$ determines the optimal resetting policy. By definition, when $\theta\in \Omega(t)$, a reorientation would increase the total cost. Hence, the agent should reorient its direction only if $\theta\notin\Omega(t)$. Since $\Omega(t)$ is dynamically coupled to the solution $J(\theta,t)$, this is a free-boundary problem for \eqref{eq:belmann}, well known in the theory of parabolic (diffusion) equations as a Stefan problem \cite{crank1984free} that must in general be solved numerically.

\subsection{Optimal navigation strategy}

Using an explicit Euler  scheme \newpart{(for details, see the supplementary code \cite{supmat})}, we integrate numerically Eq.~\eqref{eq:belmann}, updating the domain $\Omega(t)$ at each timestep. The resulting domain $\Omega(t)$ is shown in Fig.~\ref{fig:omegat}a. We find $\Omega(t)=(-\theta_a,\theta_a)$, implying that optimal reorientations are triggered when the angle $|\theta| > \theta_a(z)$. Close to the final time $t_f$, we find $\Omega(t)=[-\pi,\pi]$, meaning that when little time is left, the benefits of a reorientation would no longer outweigh its cost. Indeed, the direction should only be actively controlled for $D(t_f-t)\!>\!\log\left(2/(2-z)\right)$ and $z<z_c=2$ \cite{supmat}. For $z\geq z_c$, the cost is so high that the optimal policy is to let the system evolve freely without reorientations, independently of $t$. On the other hand, when $D(t_f-t)\gg 1$ and $z<z_c$, the optimal policy becomes time-independent, and the problem can be studied analytically.

In this stationary regime $D(t_f-t)\gg 1$, we have $\Omega=[-\theta_a,\theta_a]$, where $\theta_a$ is constant, and the optimal cost-to-go increases linearly in time. Plugging the ansatz $J(\theta,t)=j(\theta)+v^* t$, where $v^*$ describes the optimal speed of the agent, into Eq.~\eqref{eq:belmann}, we get
\begin{equation}
-v^*=D j''(\theta)-v_0\cos(\theta)\,,\quad\quad\theta\in\Omega\,,\\ 
\end{equation}
where  $\Omega=\{\theta:\,j(\theta)\leq j(0)+x_c\}$. The most general solution reads
\begin{equation}
    j(\theta)=-\frac{v^*}{D}\frac{\theta^2}{2}+a\theta+b-\frac{v_0}{D}\cos(\theta)\,.
\end{equation}
Imposing the boundary conditions
\begin{equation}
    j(\pm\theta_a)=j(0)+x_c\,,j'(\pm\theta_a)=0\,,
\end{equation}
we find 
\begin{equation}
    \frac12 \sin(\theta_a)\theta_a+\cos(\theta_a)=1-z\,, \label{eq:thetaa}
\end{equation}
and
\begin{equation}
    v^*=v_0 \frac{\sin(\theta_a)}{\theta_a}\,.
    \label{eq:vstar}
\end{equation}
 In Figs.~\ref{fig:omegat}b and \ref{fig:omegat}c, we show the optimal speed $v^*$ and activation angle  $\theta_a$, in excellent agreement with numerical simulations of Eq.~\eqref{eq:model_0} (see \cite{supmat} for details on the simulations). The threshold angle $\theta_a(z)$ increases with increasing noise levels (or, equivalently, increasing cost of resetting); when the noise is low, the optimal threshold is positioned closely to the preferred angle, indicating a more stringent control over the system. In fact, for $z\to 0$, we find the activation angle vanishes as $\theta_a\approx (24z)^{1/4}$ and the maximal velocity $v_0$ is approached as $v^*\approx v_0(1-\sqrt{2z/3})$. Furthermore, for $z>z_c$, we find $v^*=0$. \newpart{In Fig.~\ref{fig:simulations_trajectories}, we show optimally controlled trajectories for different values of the environmental noise $D$. As predicted, the effective displacement in the preferred ($x$) direction decreases with increasing noise levels. On the other hand, the lateral displacement stays roughly constant.}
 

\newpart{In the Appendix \cite{supmat}, we compare the optimal reorientation strategy with a Poissonian strategy, where correction events occur at random times following a Poisson process with rate $r_{\rm corr}$. Using the asymptotic results from Ref.~\cite{kumar2020active}, we analytically demonstrate that our strategy consistently outperforms the Poissonian approach, even when the rate $r_{\rm corr}$ is optimized. These findings are further validated through numerical simulations presented in the Appendix \cite{supmat}.}


\begin{figure}
    \centering
    \includegraphics[scale=0.8]{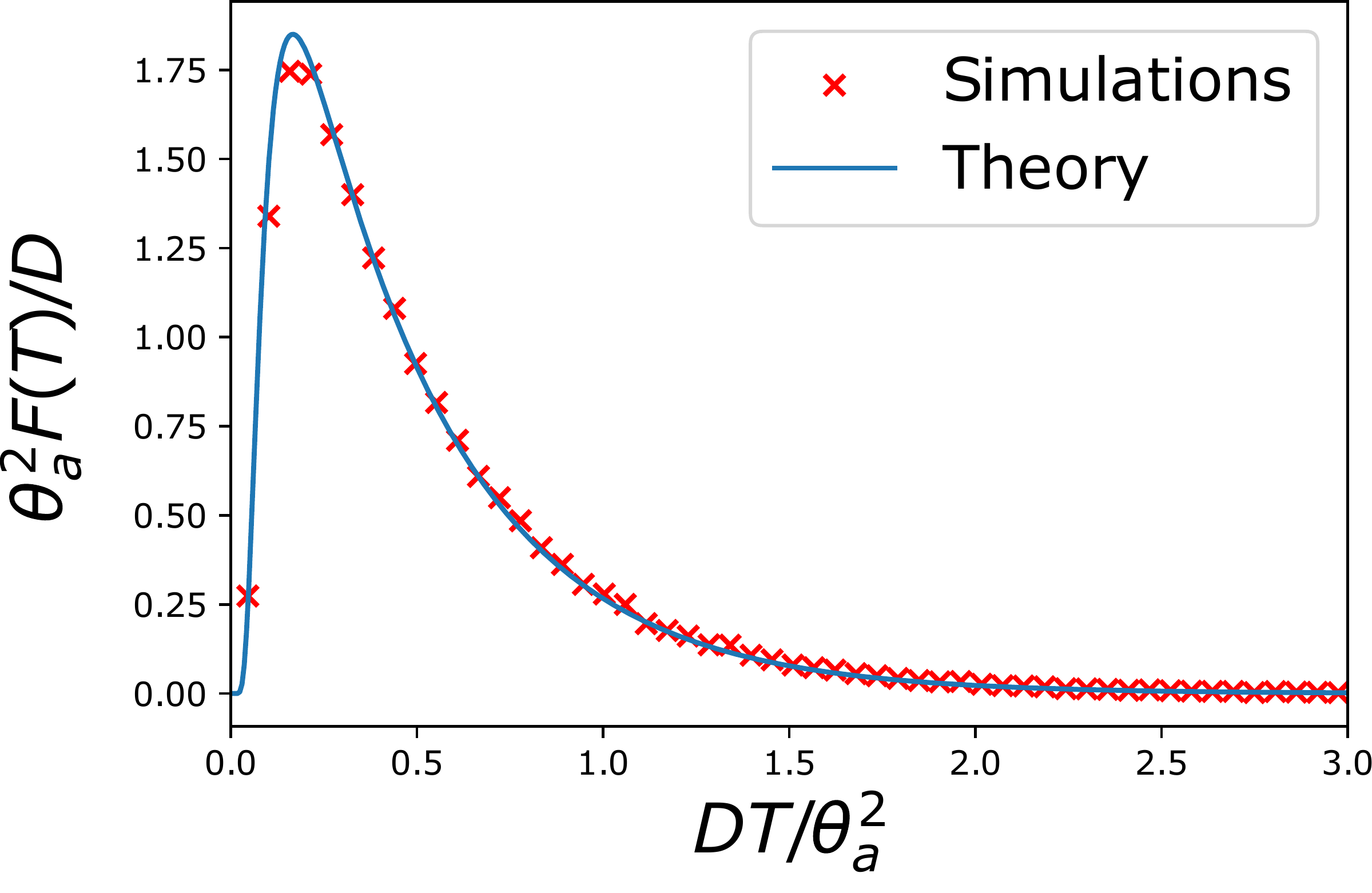}
    \caption{The scaled probability density function $\theta_a^2F(T)/D$ plotted as a function the scaled time $DT/\theta_a^2$ between two reorientations. The blue line shows the formula in Eq.~\eqref{eq:fpt}, while the red crosses are from Langevin numerical simulations of Eq.~\ref{eq:model_0} with $10^7$ samples. \label{fig:mfpt}}
\end{figure}

When $D t\gg1$ and $D(t_f-t)\gg1$, the angle $\theta$ reaches a steady state with probability density function (PDF) $P_{\rm ss}(\theta)$. Adopting the first-passage resetting techniques ~\cite{de2020optimization,de2021optimization}, we find \cite{supmat}
\begin{equation}
    P_{\rm ss}(\theta)=\frac{\theta_a-|\theta|}{\theta_a^2}\,,\quad |\theta|<\theta_a\,.
\end{equation}
Thus, the PDF of $\theta$ is maximal at the preferred direction $\theta=0$ and decays linearly away from it. Similarly, defining $F(T)$ as the PDF of the time $T$ between two orientation events, we find the following scaling form \cite{supmat,redner2001guide},  
\begin{eqnarray}
     F(T)&\!=\!&\frac{D}{\theta_a^2}f\left(\frac{DT}{\theta_a^2}\right), {\rm ~~where}\nonumber \\
     f(w)&\!=\!&\pi\sum_{n=0}^{\infty}(-1)^n (2n+1) e^{-\pi^2(2n+1)^2w/4}
\,.\label{eq:fpt}
\end{eqnarray}
This function $f(w)$ is non-monotonic, with a maximum at $w\approx 0.17$ (see Fig.~\ref{fig:mfpt}),  an essential singularity $f(w)\!\approx\! w^{-3/2}e^{-1/(4w)}/\sqrt{\pi}$ as $w\to 0$, and decays exponentially $f(w)\!\approx\! \pi e^{-\pi^2w/4}$ for $w\to\infty$. The mean time between correction events is given by \cite{supmat}
\begin{eqnarray}
    \langle T\rangle= \frac{\theta_a^2}{2D}\,.\label{eq:avg_T}
\end{eqnarray}
We note that even though the activation angle $\theta_a$ increases with increasing noise via Eq.~\eqref{eq:thetaa}, the mean time $\langle T\rangle$ is a decreasing function of $D$, in agreement with the empirical observations that reorientations are more frequent in rougher environments \cite{baird2012dung}. For small $D$, we find $\langle T\rangle\sim 1/\sqrt{D}$, in agreement with \cite{peleg2016optimal}.

\subsection{Additional sources of uncertainty}

So far, we have considered the ideal case where the agent has perfect knowledge and control of its current direction, $\theta(t)$. However, various sources of uncertainty can impact the precision and speed of dung beetles \cite{baird2012dung,khaldy2019effect,dacke2021dung}. Experiments show that the angle at which reorientations occur is not constant, suggesting uncertainty in sensory acquisition \cite{baird2012dung}. Indeed, even if beetles have good knowledge of their direction after a reorientation, this knowledge gradually deteriorates over time due to imperfect sensory data, execution errors, and limitations in memory.

In this section, we consider the effect of these additional sources of noise on both strategy and performance. We address errors associated with both motor execution and perception. In the latter case, we distinguish between two types of perception errors: noisy measurements, where the perceived signal is corrupted by noise, and partial observability, where the agent has access to only some of the relevant information needed to determine its direction. Both execution and perception errors are known to influence the performance of dung beetles \cite{khaldy2019effect}. For a schematic representation of the different types of uncertainty considered, see Fig.~\ref{fig:v_star_noise}.

In experiments, both perception and motor errors can be externally controlled \cite{dacke2021dung}. For instance, motor errors can be increased by performing experiments in rougher environments, either natural or artificially generated \cite{baird2012dung}. Similarly, perception errors can be enhanced by altering the location or visibility of external visual cues \cite{khaldy2019effect,khaldy2020straight}.

\subsubsection{Execution errors}

 To incorporate execution errors, we assume that the position of the agent evolves according to 
\begin{equation}
    \dot{x}(t)=v_0\cos[\theta(t)+\epsilon(t)]\,,
    \dot{y}(t)=v_0\sin[\theta(t)+\epsilon(t)]\,,
\end{equation}
where the dynamics of $\theta(t)$ is unchanged and the execution error $\epsilon(t)$ is a uniform random variable in $[-\delta,\delta]$, \newpart{with time correlation $\tau\gg1/D$} (see Fig.~\ref{fig:v_star_noise}a and \cite{supmat} for details). Adapting the optimal control equation \eqref{eq:belmann} to investigate this model \cite{supmat}, we find that the angle $\theta_a$ satisfies the modified condition
\begin{equation}
    \frac12 \sin(\theta_a)\theta_a+\cos(\theta_a)=1-z\frac{\delta}{\sin(\delta)}\,. \label{eq:thetaa_delta}
\end{equation}
We see that the additional noise effectively increases the cost of resetting by a factor $\delta/\sin(\delta)\geq 1$; as $\delta\to 0$, we recover Eq.~\eqref{eq:thetaa}. In this case, the optimal velocity reads \cite{supmat}
\begin{equation}
    v^*=v_0\frac{\sin(\delta)}{\delta}\frac{\sin(\theta_a)}{\theta_a}\,.\label{v_delta}
\end{equation}
Interestingly, even in the limit of low environmental noise $z\to 0$, the optimal speed $v^*=v_0 \sin(\delta)/\delta\leq v_0$ is bounded away from the maximal value $v_0$ (see Fig.~\ref{fig:v_star_noise}a). 

\begin{figure*}
    \centering
    \includegraphics[width=\linewidth]{images_main_text_png/Figure5.png}
    \caption{Optimal scaled speed $v^*/v_0$ as a function of the scaled resetting parameter $z=x_cD/v_0$ under three different noise conditions. In all schematic representations above the graphs, the red arrow indicates the direction of the agent within the $(x,y)$ plane, and the thought balloons indicate the observed angle. \textbf{Effect of execution errors} (panel {\bf a}): The agent observes the direction $\theta$, which is corrupted by the execution error $\epsilon$, uniformly distributed in the interval $[-\delta,\delta]$ and uncorrelated in time. The angle $\theta$ evolves as a Brownian motion with diffusion constant $D$. Different curves in the graph correspond to different values of $\delta$, describing the magnitude of the execution noise. \textbf{Effect of partial observations} (panel {\bf b}): The agent can only observe a portion $\theta_1$ of the direction $\theta=\theta_1+\theta_2$. The two angles $\theta_1$ and $\theta_2$ evolve independently as Brownian motions with diffusion constants $D_1$ and $D_2$. Different curves in the graph correspond to different values of the observability ratio $\ell=D_1/D_2$. \textbf{Effect of noisy observations} (panel {\bf c}): The agent observes a noisy version $\theta+\theta_n$ of the real direction $\theta$. The angles $\theta$ and $\theta_n$ evolve independently as Brownian motions with diffusion constants $D$ and $D_n$. Different curves in the graph correspond to different values of the signal-to-noise ratio $r=D/D_n$. In each case, a reorientation consists in setting the angular variables $\theta$, $\theta_1$, $\theta_2$, or $\theta_n$ to zero. The crosses are the results of Langevin simulations with $Dt_f=10^4$.
    \label{fig:v_star_noise}}
\end{figure*}

\subsubsection{Partial observability}

To account for error accumulation, we consider the case when the agent has only partial knowledge of its current direction (see Fig.~\ref{fig:v_star_noise}b). As an illustrative example, consider the case when the agent's sensors could only discern the presence of terrain irregularities, while other factors influencing the dynamics of $\theta$ remain unobservable. We consider the dynamics in Eq.~\eqref{eq:model_0} but with $\theta(t)$ defined as the sum of two components: $\theta_1(t)$, the observable angle, and $\theta_2(t)$, the hidden angular deviation. The two angles evolve as $\dot{\theta_i}(t)=\eta_i(t)$, where $\langle\eta_i(t)\eta_j(t')\rangle=2D_i\delta_{ij}\delta(t-t')$. We assume that when the agent reorients its direction, both $\theta_1(t)$ and $\theta_2(t)$ are reset to zero. Hence, the knowledge of $\theta(t)$ is perfect after a reset but degrades over time as $\theta_2(t)$ grows.

Since the agent can only make decisions based on the current value of $\theta_1$, we consider the control policy of resetting the direction when $|\theta_1|\geq \theta_a$. For a fixed value of $\theta_a$, we compute the steady-state PDF $P_{\rm ss}(\theta)$ of $\theta=\theta_1+\theta_2$. Calculating $P_{\rm ss}(\theta)$ (see \cite{supmat} for details), we find
\begin{equation}
     P_{ss}(\theta)=\frac{1}{\pi\theta_a^2}\int_{-\infty}^{\infty}dk~e^{-ik\theta}~\frac{1-\frac{\cos(k\theta_a)}{\cosh\left(k\theta_a/\sqrt{\ell}\right)}}{(1+1/\ell)k^2}\,,
     \label{eq:pss-noise}
\end{equation}
where $\ell=D_1/D_2$ is the ratio of the diffusivities of the observable and the hidden angular variables. The velocity in the steady state is then given by
\begin{align}
    v&\equiv\lim_{t_f\to\infty}\frac{1}{t_f}\left[ -\int_{0}^{t_f}dt~v_0\cos(\theta(t))+x_c N(t_f)\right] \nonumber\\
    &= -v_0\langle \cos(\theta)\rangle_{\rm ss}+x_c/\langle T\rangle\,, 
\end{align}
where $\langle \cdot\rangle_{\rm ss}$ indicates the steady state average in Eq.~\eqref{eq:pss-noise}. The average time between reorientations is given by $\langle T\rangle=\theta_a^2/(2D_1)$. Using Eq.~\eqref{eq:pss-noise}, we find
\begin{equation}
    v=\frac{2v_0}{\theta_a^2(1+1/\ell)}\left[1-\frac{\cos(\theta_a)}{\cosh\left(\theta_a/\sqrt{\ell}\right)}\right]-\frac{2v_0z}{\theta_a^2}\,, \label{v_sensory}
\end{equation}
where $z=D_1x_c/v_0$ is the scaled resetting cost. By minimizing this expression as a function of $\theta_a$, we find the optimal speed $v^*$ \cite{supmat}. Fig.~\ref{fig:v_star_noise}b shows that $v^*$ increases with $\ell$ for all values of $z$, implying that higher observability leads to higher speed \cite{griffith1971observability}. In the limit $\ell\to\infty$ we recover our original model without measurement errors. Interestingly, at variance with the case of execution errors, the maximal velocity $v_0$ can be reached in the low-noise limit $z\to 0$ for all values $\ell$. In particular, for $z\to0$, we find the following asymptotic behaviors
\begin{equation}
    \theta_a\approx\left(\frac{24\ell}{5+\ell}\right)^{1/4}z^{1/4}\,,\quad
    \frac{v^*}{v_0}\approx 1-\sqrt{\frac{2(5+\ell)z}{3\ell}}\,.
\end{equation}
\newpart{We observe that for both quantities the scaling with the environmental noise $z$ is the same as in the case of perfect observability but with an $\ell$-dependent prefactor. In particular, we find that for increasing observability $\ell$, the optimal control strategy becomes less stringent (larger $\theta_a$) and the speed increases.}

\newpart{To find the maximal environmental noise $z_c(\ell)$ that the agent can tolerate with a finite net velocity, we simply have to set the activation angle to its maximal value $\theta_a=\pi$ and the velocity to zero $v=0$ in Eq.~\eqref{v_sensory}, yielding
\begin{eqnarray}
    z_c(\ell)=\frac{\ell}{1+\ell}\left(1+\frac{1}{\cosh(\pi/\sqrt{\ell})}
\right)\,.
\end{eqnarray} 
We recall that for $z<z_c(\ell)$ the agent is able to identify an optimal strategy with nonzero velocity while for $z>z_c(\ell)$ the noise is too high and the optimal strategy is to never perform reorientations, resulting in $v=0$ on average. In the limit of large $\ell$, we recover the result derived in the previous sections $z(\ell)\approx 2-(4+\pi^2)/(2\ell)$, while for small observability ($\ell\to 0$) the maximal tolerable noise vanishes as $z_c(\ell)\approx \ell$.}

\subsubsection{Measurement errors}
As another variant of noisy sensory acquisition we assume that the agent can observe all relevant factors that affect its direction ($\ell\to \infty$), but that these observations are acquired through noisy sensory channels. Thus, although the real direction $\theta$ of the agent evolves as a Brownian motion with diffusion constant $D$, the observed angle is $\theta_1(t)=\theta(t)+\theta_n(t)$, where $\theta_n(t)$ describes the accumulated measurement error. We assume that $\theta_n(t)$ independently evolves as a Brownian motion with diffusion constant $D_n$ and that after a reorientation both $\theta$ and $\theta_n$ are set to zero. The control strategy is again to reorient the direction if $|\theta_1|>\theta_a$, leading to the speed (see \cite{supmat} for the derivation)
\begin{equation}
   v=\frac{2v_0}{\theta_a^2}(1+1/r)\left[1-z-\frac{\cos\left(\frac{\theta_a}{1+1/r}\right)}{\cosh\left(\frac{\theta_a\sqrt{1/r}}{1+1/r}\right)}\right]\,,\label{eq:v_meas_err}
\end{equation}
where $z=Dx_c/v_0$ and $r=D/D_n$ is the signal-to-noise ratio. In the limit $r\to\infty$, we recover the results of our original model without sensory noise. In this case, the optimal speed can be obtained by minimizing Eq.~\eqref{eq:v_meas_err} with respect to $\theta_a\in [0,\pi]$, see Fig.~\ref{fig:v_star_noise}c. The qualitative behavior is similar to the previous case.  The optimal speed increases with the signal-to-noise ratio and the maximal speed $v_0$ is attained for all values of $r$ for $z\to0$, when
\begin{equation}
\theta_a\approx\frac{2^{3/4}3^{1/4}(1+r)^{3/4}}{r^{1/2}(5+r)^{1/4}}z^{1/4}\,,\quad
    \frac{v^*}{v_0}\approx1-\sqrt{\frac{2r^2(r+5)z}{3r^2(r+1)}}\,.
\end{equation}
\newpart{As in the case of partial observability, the scaling of both activation angle and velocity with $z$ is unchanged, but with model-dependent prefactors. Interestingly, the prefactor in the asymptotic expression for $\theta_a$ has a non-monotonic behavior with $r$, with a minimum at intermediate values. As expected, the optimal speed however always increases with the signal-to-noise ratio $r$.}

\newpart{In the case of measurement noise, the maximal environmental noise $z_c(r)$ beyond which the velocity vanishes reads
\begin{eqnarray}
 z_c(r)=1-\frac{\cos(\pi r/(1+r))}{\cosh(\pi \sqrt{r}/(1+r))} \,. 
\end{eqnarray}
This quantity vanishes in the absence of signal ($r\to 0$) as $z_c(r)\approx \pi^2 r/2$. In the absence of measurement errors ($r\to \infty$), it approaches the asymptotic value $z_c=2$ as $z_c(r)\approx 2-\pi^2/(2r)$.}

\subsection{Non-instantaneous reorientations}

When a beetle reorients its direction, a finite amount of time is required to operate such correction. So far, we have assumed that the reorientations are instantaneous and we have described this time delay with the effective cost $x_c$. Here, we show that our results can be extended to a model where the correction mechanism is explicitly described. This model will allow us to determined the fraction of the total time spent in the reorientation phase as a function of the noise level. 

We assume that the agent can switch between two navigation modes: an attention mode and a speed mode. In the first, it moves at low speed $v_1$ but the direction $\theta(t)$ evolves according to
\begin{equation}
    \dot{\theta}(t)=-a\operatorname{sign}(\theta)+\eta(t)\,, \label{eq:drift}
\end{equation}
where $a>0$ is the magnitude of the correcting drift and $\eta(t)$ is as before Gaussian white noise with rotational diffusion constant $D$. The drift continuously steers the angle $\theta$ towards the preferred direction. In the second mode, the agent moves at higher speed $v_0>v_1$ but has no control over its direction, which evolves as $\dot{\theta}(t)\!=\!\eta(t)$. We do not consider the cost of switching from one mode to the other. The goal is to maximize the displacement $\langle x(t_f)\rangle$.


\begin{figure}
    \centering
    \includegraphics[width=\linewidth]{images_main_text_png/Figure6.png}
    \caption{ {\bf a)} Optimal activation angle $\theta_a/\pi$, {\bf b)} speed $v^*/v_0$, and {\bf c)} franction $p_r$ of the total time spent in the reorientation phase as a function of the inverse Péclet number $1/\Pe=D/a$ in the case of non-instantaneous reorientations with $v_1 = 0$. The red crosses are the results of numerical simulations with $Dt_f=10^4$. 
    \label{fig:non_inst} }
\end{figure}

The optimal switching strategy can be derived by adapting the Bellman equation \eqref{eq:belmann} to the modified dynamics. However, to investigate the large time-horizon limit ($t_f- t\gg D ^{-1}$), it is sufficient to study the stationary properties of the dynamics of $\theta$. The optimal strategy is to move with high speed only if $|\theta|<\theta_a$, where $\theta_a$ has to be chosen to minimize the cost function. Interestingly, the dynamics of $\theta$ can be rewritten as the equilibrium Langevin equation
\begin{equation}
    \dot{\theta}(t)=-\partial_{\theta}V(\theta)+\eta(t)\,,
\end{equation}
where the potential $V(\theta)=a(|\theta|-\theta_a)$ for $|\theta|>\theta_a$ and $V(\theta)=0$ otherwise. As a consequence, the stationary state of $\theta$ is given by the equilibrium Boltzmann measure $P_{\rm eq}(\theta)=\exp[-V(\theta)/D]/Z$, where $Z$ is the partition function ensuring normalization. By defining $v(\theta)=v_0$ if $|\theta|<\theta_a$ and $v(\theta)=v_1$ otherwise, we can write the average speed as
\begin{equation}
    v=\lim_{t_f\to\infty}\frac{1}{t_f}\int_{0}^{t_f}dt~v(\theta(t))\cos(\theta(t))=\langle v(\theta)\cos(\theta)\rangle_{\rm eq}\,,
\end{equation}
where $\langle\cdot\rangle_{\rm eq}$ indicates an average with respect to $P_{\rm eq}(\theta)$. In the case $v_1=0$, we find
\begin{equation}
    v=v_0\frac{\sin(\theta_a)}{\theta_a+\left(1-e^{-{\rm Pe}(\pi-\theta_a)}\right)/{\rm Pe}}\,,\label{eq:speed_nonist}
\end{equation}
where the Péclet number ${\rm Pe}=a/D$ is the ratio of the drift to the diffusion. By minimizing this expression with respect to $\theta_a$, we find the optimal speed $v^*$ as a function of ${\rm Pe}$, see Figs.~\ref{fig:non_inst}a and \ref{fig:non_inst}b.

The low-noise behavior (${\rm Pe}\gg 1$) is qualitatively similar to the previous cases. We find that the optimal threshold $\theta_a$ approaches zero as $\theta_a\approx ({\rm Pe}/3)^{-1/3}$ and that the maximal speed $v_0$ is approached as $v^*\!\approx\! v_0 -v_0({\rm Pe}/3)^{-2/3}/2$. In the high-noise regime ${\rm Pe} \ll 1$, the angle $\theta_a$ approaches $\pi/2$ as $\theta_a\!\approx\! \pi/2-2 \Pe$. Indeed, when $|\theta|>\pi/2$, the agent is moving in the wrong direction and it is therefore convenient to switch to the slow state, regardless of the noise level. The resulting speed behaves as $v^*\!\approx\!v_0(1/\pi+\Pe/8)$. The asymptotic value $v_0/\pi$ can be explained by the fact that, in the high-noise limit, the equilibrium PDF of $\theta$ approaches a uniform distribution over $[-\pi,\pi]$.

As non-instantaneous reorientation are explicitly described, the current model allows to compute another key quantity: the fraction $p_r$ of the time that the agent spends reorienting its direction. This quantity can be easily measured in experiments and could hence provide a clear prediction on the relation between reorientation time and environmental noise. Since the agent is in the reorientation phase when $|\theta|>\theta_a$, this fraction can be readily computed as 
\begin{eqnarray}
    p_r=\int_{-\pi}^{\pi}d\theta~P_{\rm eq}(\theta) H(|\theta|-\theta_a)\,,
\end{eqnarray}
where $H(z)$ is the Heaviside step function, with $H(z)=1$ for $z>0$ and $H(z)=0$ otherwise. Performing the integral, we find
\begin{eqnarray}
    p_r=1-\frac{\Pe~ \theta_a}{1+\Pe~ \theta_a -e^{-\Pe(\pi-\theta_a)}}\,.\label{eq:pr}
\end{eqnarray}
Note that here $\theta_a$ is chosen to maximize the speed in Eq.~\eqref{eq:speed_nonist} and is, therefore, a function of $\Pe$.

The reorientation fraction $p_r$ is shown in Fig.~\ref{fig:non_inst}c) and is in perfect agreement with numerical simulations. Interestingly, $p_r$ turns out to be a decreasing function of the Péclet number $\Pe$. This result is in agreement with the empirical observation that in more noisy terrains a greater amount of time is devoted to correct the direction, due to more frequent reorientations \cite{baird2012dung}. In particular, in the low-noise regime $\Pe\gg1$, the fraction of the total time spent in reorienting the direction goes to zero as  $p_r\approx 3^{-1/3} \Pe^{-2/3}$. On the other hand, in the high-noise limit $\Pe\ll1$, approximately half of the total time is devoted to reorientations.

\section{Discussion}

Inspired by the intermittent navigational strategy of the dung beetle, we have defined a minimal model that describes the alternation of phases of movement and phases of directional correction. Using optimal control theory, we have identified an optimal strategy that integrates egocentric and geocentric schemes to maximize the speed in a given direction. We have extended our result to account for both sensing and actuation errors. Our results lead to a range of testable predictions:
\begin{enumerate}
    \item The relation between the time $T$ between reorientations and noise strength, given in Eq.~\eqref{eq:avg_T}. Motor noise intensity can be experimentally varied by conducting experiments on natural or artificial terrains of varying roughness \cite{baird2012dung}.
    \item The effective speed $v^*$ as a function of errors in measurement and actuation, given Eqs.~\eqref{eq:vstar} for environmental noise, \eqref{v_delta} for execution errors, \eqref{v_sensory} for partial observability, and \eqref{eq:v_meas_err} for measurement errors (see also Fig.~\ref{fig:v_star_noise} for a schematic representation of the different settings). Perception errors can be manipulated by changing the position or visibility of visual cues, such as the sun for outdoor experiments or LED lights in laboratory settings \cite{khaldy2019effect,khaldy2020straight}. For instance, the sun position can be apparently displaced using mirrors and its intensity can be modulated by conducting experiments under an overcast sky \cite{dacke2021dung}.
    \item The fraction $p_r$ of time spent in the reorienting phase as a function of environmental noise (Eq.~\eqref{eq:pr}). Analogously to (i), this prediction could be verified by conducting the same experiment in several landscapes of variable roughness.
\end{enumerate}
Since motor and perception errors can be controlled with existing experimental techniques \cite{baird2012dung,khaldy2019effect,dacke2021dung}, an immediate question that this raises is that of comparing our results to experimental data on the navigation of dung beetles \cite{foster2021light}. This would involve measuring the three main quantities $T$, $v^*$, and $p_r$, that can be directly obtained from recorded two-dimensional trajectories of the dung beetles rolling behavior.

More broadly, our approach can also be used in general search situations where the cues about the target location can be obtained from different sensory channels. Indeed this scenario of switching between two modes associated with different modalities of information retrieval and action is quite common, e.g. dogs sniff the ground to get accurate local information and also lift their heads into the boundary layer to get noisy long-range information \cite{thesen1993behaviour}, while moths alternate between casting and tracking \cite{reddy2022olfactory,rigolli2022alternation}. \newpart{Additionally, our framework could be extended to account for scenarios where the reorientation angle is not reset to exactly zero but instead follows a distribution. This would allow us to explore the impact of residual deviations on the overall navigation strategy and performance. Another intriguing extension of our framework would involve studying optimal resetting strategies and trajectories in systems such as cellular protrusions in zebrafish, which are thought to find the optimal balance between ballistic and diffusive searches to locate targets, as described in \cite{park2022zebrafish}.}  Finally, our model is but a first step in exploring the role of measurement, estimation, representation, sensory integration known to be important in the context of navigation \cite{haberkern2016studying,kim2019generation}.  

While we have focused on using natural systems to motivate our question, there are artificial systems that have recently been studied from the perspective of control. In a series of recent works \cite{mano2017optimal,selmke2018theory,selmke2018theory2}, a novel experimental technique has been introduced to control the direction of Janus particles to induce material transport. Modulating an external electric field, a controller can align the particle heading to a preferred direction, and a natural question is to ask if the framework of our paper can be tested in such artificial manifestations as well.\\

\enlargethispage{20pt}

{\bf Author Contributions} F.M. developed the mathematical model, performed analytical calculations and numerical simulations, and drafted the manuscript. L.M. conceived the project, contributed to the design of the mathematical framework, supervised the work, and contributed to the writing and revision of the manuscript.

{\bf Data accessibility.} This study does not use experimental data. The code used to numerically integrate the optimal control equations, perform numerical simulations and generate figures is available at \url{https://doi.org/10.5281/zenodo.14773662} or \url{https://github.com/francescomori/optimal_reorientations}

{\bf Funding statement.} This work was supported by a Leverhulme Trust International Professorship Grant (No. LIP-2020-014), the Simons Foundation and the Henri Seydoux Fund.


\vskip2pc

\printbibliography

@misc{supmat,
  note = {See Supplemental Material for additional details on the analytical derivations. The code used to numerically integrate the optimal control equations, perform numerical simulations and generate figures is available at \url{https://doi.org/10.5281/zenodo.14773662} or \url{https://github.com/francescomori/optimal_reorientations}.},
}

@misc{image,
note={\href{https://commons.wikimedia.org/wiki/File:Dung_beetle_(12593432075).jpg}{Dung beetle (12593432075)} and  \href{https://commons.wikimedia.org/wiki/File:Dung_beetle_(12593865234).jpg}{Dung beetle (12593865234)} by \href{https://www.flickr.com/people/21162417@N07}{flowcomm}. CC BY 2.0}
}

@article{freas2022basis,
  title={The basis of navigation across species},
  author={Freas, Cody A and Cheng, Ken},
  journal={Annu. Rev. Psychol.},
  volume={73},
  pages={217--241},
  year={2022},
  publisher={Annual Reviews}
}

@article{wiltschko2023animal,
  title={Animal navigation: how animals use environmental factors to find their way},
  author={Wiltschko, Roswitha and Wiltschko, Wolfgang},
  journal={Eur Phys J Spec Top},
  volume={232},
  number={2},
  pages={237--252},
  year={2023},
  publisher={Springer}
}

@article{menti2023towards,
  title={Towards a unified vision on animal navigation},
  author={Menti, Giulio Maria and Meda, Nicola and Zordan, Mauro A and Megighian, Aram},
  journal={Eur. J. Neurosci.},
  volume={57},
  number={12},
  pages={1980--1997},
  year={2023},
  publisher={Wiley Online Library}
}

@article{wehner1996visual,
  title={Visual navigation in insects: coupling of egocentric and geocentric information},
  author={Wehner, R{\"u}diger and Michel, Barbara and Antonsen, Per},
  journal={J. Exp. Biol.},
  volume={199},
  number={1},
  pages={129--140},
  year={1996},
  publisher={The Company of Biologists Ltd}
}

@article{peleg2016optimal,
  title={Optimal switching between geocentric and egocentric strategies in navigation},
  author={Peleg, O and Mahadevan, L},
  journal={Royal Soc. Open Sci.},
  volume={3},
  number={7},
  pages={160128},
  year={2016},
  publisher={The Royal Society}
}

@article{yin2020simulating,
  title={Simulating rolling paths and reorientation behavior of ball-rolling dung beetles},
  author={Yin, Zhanyuan and Zinn-Bj{\"o}rkman, Leif},
  journal={J. Theor. Biol.},
  volume={486},
  pages={110106},
  year={2020},
  publisher={Elsevier}
}

@book{byrne2011visual,
  title={The visual ecology of dung beetles, in L. W. Simmons \& T. J. Ridsill-Smith (Eds.), Ecology and evolution of dung beetles},
  author={Byrne, Marcus and Dacke, Marie},
  pages={177--199},
  year={2011},
  publisher={ John Wiley \& Sons, Chichester, U.K.}
}

@article{baird2012dung,
  title={The dung beetle dance: an orientation behaviour?},
  author={Baird, Emily and Byrne, Marcus J and Smolka, Jochen and Warrant, Eric J and Dacke, Marie},
  journal={PLoS One},
  volume={7},
  number={1},
  pages={e30211},
  year={2012},
  publisher={Public Library of Science San Francisco, USA}
}

@book{bellman1965dynamic,
  title={Dynamic programming and modern control theory},
  author={Bellman, Richard and Kalaba},
  volume={81},
  year={1965},
  publisher={Academic Press, New York}
}

@article{de2023resetting,
  title={Resetting in stochastic optimal control},
  author={De Bruyne, Benjamin and Mori, Francesco},
  journal={Phys. Rev. Research},
  volume={5},
  number={1},
  pages={013122},
  year={2023},
  publisher={APS}
}

@article{basu2018active,
  title={Active Brownian motion in two dimensions},
  author={Basu, Urna and Majumdar, Satya N and Rosso, Alberto and Schehr, Gr{\'e}gory},
  journal={Phys. Rev. E},
  volume={98},
  number={6},
  pages={062121},
  year={2018},
  publisher={APS}
}

@article{evans2011diffusion,
  title={Diffusion with stochastic resetting},
  author={Evans, Martin R and Majumdar, Satya N},
  journal={Phys. Rev. Lett.},
  volume={106},
  number={16},
  pages={160601},
  year={2011},
  publisher={APS}
}

@article{evans2011diffusionoptimal,
  title={Diffusion with optimal resetting},
  author={Evans, Martin R and Majumdar, Satya N},
  journal={J. Phys. A Math. Theor.},
  volume={44},
  number={43},
  pages={435001},
  year={2011},
  publisher={IOP Publishing}
}

@article{kumar2020active,
  title={Active Brownian motion in two dimensions under stochastic resetting},
  author={Kumar, Vijay and Sadekar, Onkar and Basu, Urna},
  journal={Phys. Rev. E},
  volume={102},
  number={5},
  pages={052129},
  year={2020},
  publisher={APS}
}

@article{abdoli2021stochastic,
  title={Stochastic resetting of active Brownian particles with Lorentz force},
  author={Abdoli, Iman and Sharma, Abhinav},
  journal={Soft Matter},
  volume={17},
  number={5},
  pages={1307--1316},
  year={2021},
  publisher={Royal Society of Chemistry}
}

@article{sar2023resetting,
  title={Resetting mediated navigation of active Brownian searcher in a homogeneous topography},
  author={Sar, Gourab Kumar and Ray, Arnob and Ghosh, Dibakar and Hens, Chittaranjan and Pal, Arnab},
  journal={Soft Matter},
  year={2023},
volume={19},
  pages={4502-4518},
  publisher={Royal Society of Chemistry}
}

@book{crank1984free,
  title={Free and moving boundary problems},
  author={Crank, John},
  publisher={Oxford University Press, New York},
  year={1984}
}

@article{de2020optimization,
  title={Optimization in first-passage resetting},
  author={De Bruyne, B and Randon-Furling, J and Redner, S},
  journal={Phys. Rev. Lett.},
  volume={125},
  number={5},
  pages={050602},
  year={2020},
  publisher={APS}
}

@article{de2021optimization,
  title={Optimization and growth in first-passage resetting},
  author={De Bruyne, B and Randon-Furling, Julien and Redner, S},
  journal={J. Stat. Mech. Theory Exp.},
  volume={2021},
  number={1},
  pages={013203},
  year={2021},
  publisher={IOP Publishing}
}

@article{evans2020stochastic,
  title={Stochastic resetting and applications},
  author={Evans, Martin R and Majumdar, Satya N and Schehr, Gr{\'e}gory},
  journal={J. Phys. A Math. Theor.},
  volume={53},
  number={19},
  pages={193001},
  year={2020},
  publisher={IOP Publishing}
}

@book{redner2001guide,
  title={A guide to first-passage processes},
  author={Redner, Sidney},
  year={2001},
  publisher={Cambridge university press}
}

@article{foster2021light,
  title={Light pollution forces a change in dung beetle orientation behavior},
  author={Foster, James J and Tocco, Claudia and Smolka, Jochen and Khaldy, Lana and Baird, Emily and Byrne, Marcus J and Nilsson, Dan-Eric and Dacke, Marie},
  journal={Curr. Biol.},
  volume={31},
  number={17},
  pages={3935--3942},
  year={2021},
  publisher={Elsevier}
}

@article{bijma2021sisyphus,
  title={Sisyphus and his rock: Quasi-random walk inspired by the motion of a ball transported by a dung beetle on combined terrain},
  author={Bijma, Nienke N and Filippov, Alexander E and Gorb, Stanislav N},
  journal={J. Theor. Biol.},
  volume={520},
  pages={110659},
  year={2021},
  publisher={Elsevier}
}

@article{yin2022mathematical,
  title={Mathematical modeling shows that ball-rolling dung beetles can use dances to avoid competition},
  author={Yin, Zhanyuan and Zinn-Bj{\"o}rkman, Leif},
  journal={Theor. Ecol.},
  volume={15},
  number={1},
  pages={17--28},
  year={2022},
  publisher={Springer}
}

@article{codling2008random,
  title={Random walk models in biology},
  author={Codling, Edward A and Plank, Michael J and Benhamou, Simon},
  journal={J R Soc Interface},
  volume={5},
  number={25},
  pages={813--834},
  year={2008},
  publisher={The Royal Society London}
}

@article{benichou2011intermittent,
  title={Intermittent search strategies},
  author={B{\'e}nichou, Olivier and Loverdo, Claude and Moreau, Michel and Voituriez, Raphael},
  journal={Rev. Mod. Phys.},
  volume={83},
  number={1},
  pages={81},
  year={2011},
  publisher={APS}
}

@article{cheung2007animal,
  title={Animal navigation: the difficulty of moving in a straight line},
  author={Cheung, Allen and Zhang, Shaowu and Stricker, Christian and Srinivasan, Mandyam V},
  journal={Biol. Cybern.},
  volume={97},
  pages={47--61},
  year={2007},
  publisher={Springer}
}

@misc{video,
  note = {Lund Vision Group, Dung beetle dance, video at \url{https://www.youtube.com/watch?v=w1XL711elDA}},
}

@article{bailey2018navigational,
  title={Navigational efficiency in a biased and correlated random walk model of individual animal movement},
  author={Bailey, Joseph D and Wallis, Jamie and Codling, Edward A},
  journal={Ecology},
  volume={99},
  number={1},
  pages={217--223},
  year={2018},
  publisher={Wiley Online Library}
}

@article{haberkern2016studying,
  title={Studying small brains to understand the building blocks of cognition},
  author={Haberkern, Hannah and Jayaraman, Vivek},
  journal={Curr. Opin. Neurobiol.},
  volume={37},
  pages={59--65},
  year={2016},
  publisher={Elsevier}
}

@article{park2022zebrafish,
  title={Zebrafish airinemes optimize their shape between ballistic and diffusive search},
  author={Park, Sohyeon and Kim, Hyunjoong and Wang, Yi and Eom, Dae Seok and Allard, Jun},
  journal={Elife},
  volume={11},
  pages={e75690},
  year={2022},
  publisher={eLife Sciences Publications Limited}
}

@article{kim2019generation,
  title={Generation of stable heading representations in diverse visual scenes},
  author={Kim, Sung Soo and Hermundstad, Ann M and Romani, Sandro and Abbott, LF and Jayaraman, Vivek},
  journal={Nature},
  volume={576},
  number={7785},
  pages={126--131},
  year={2019},
  publisher={Nature Publishing Group UK London}
}

@article{dacke2021dung,
  title={How dung beetles steer straight},
  author={Dacke, Marie and Baird, Emily and El Jundi, Basil and Warrant, Eric J and Byrne, Marcus},
  journal={Annu. Rev. Entomol.},
  volume={66},
  pages={243--256},
  year={2021},
  publisher={Annual Reviews}
}

@article{khaldy2019effect,
  title={The effect of step size on straight-line orientation},
  author={Khaldy, Lana and Peleg, Orit and Tocco, Claudia and Mahadevan, L and Byrne, Marcus and Dacke, Marie},
  journal={J R Soc Interface},
  volume={16},
  number={157},
  pages={20190181},
  year={2019},
  publisher={The Royal Society}
}

@article{thesen1993behaviour,
  title={Behaviour of dogs during olfactory tracking},
  author={Thesen, Aud and Steen, Johan B and D{\o}ving, Kjell B},
  journal={J. Exp. Biol.},
  volume={180},
  number={1},
  pages={247--251},
  year={1993},
  publisher={The Company of Biologists Ltd}
}

@book{viswanathan2011physics, title={The physics of foraging: an introduction to random searches and biological encounters}, author={Viswanathan, Gandhimohan M and Da Luz, Marcos GE and Raposo, Ernesto P and Stanley, H Eugene}, year={2011}, publisher={Cambridge University Press} }

@article{reddy2022olfactory,
  title={Olfactory sensing and navigation in turbulent environments},
  author={Reddy, Gautam and Murthy, Venkatesh N and Vergassola, Massimo},
  journal={Annu. Rev. Condens. Matter Phys.},
  volume={13},
  pages={191--213},
  year={2022},
  publisher={Annual Reviews}
}

@article{griffith1971observability,
  title={On the observability of nonlinear systems: I},
  author={Griffith, Ernest William and Kumar, KSP},
  journal={J. Math. Anal. Appl.},
  volume={35},
  number={1},
  pages={135--147},
  year={1971},
  publisher={Academic Press}
}

@article{rigolli2022alternation,
  title={Alternation emerges as a multi-modal strategy for turbulent odor navigation},
  author={Rigolli, Nicola and Reddy, Gautam and Seminara, Agnese and Vergassola, Massimo},
  journal={Elife},
  volume={11},
  year={2022},
  publisher={eLife Sciences Publications, Ltd}
}

@article{mano2017optimal,
  title={Optimal run-and-tumble--based transportation of a Janus particle with active steering},
  author={Mano, Tomoyuki and Delfau, Jean-Baptiste and Iwasawa, Junichiro and Sano, Masaki},
  journal={Proceedings of the National Academy of Sciences},
  volume={114},
  number={13},
  pages={E2580--E2589},
  year={2017},
  publisher={National Acad Sciences}
}

@article{selmke2018theory,
  title={Theory for controlling individual self-propelled micro-swimmers by photon nudging I: directed transport},
  author={Selmke, Markus and Khadka, Utsab and Bregulla, Andreas P and Cichos, Frank and Yang, Haw},
  journal={Physical Chemistry Chemical Physics},
  volume={20},
  number={15},
  pages={10502--10520},
  year={2018},
  publisher={Royal Society of Chemistry}
}

@article{selmke2018theory2,
  title={Theory for controlling individual self-propelled micro-swimmers by photon nudging II: confinement},
  author={Selmke, Markus and Khadka, Utsab and Bregulla, Andreas P and Cichos, Frank and Yang, Haw},
  journal={Physical Chemistry Chemical Physics},
  volume={20},
  number={15},
  pages={10521--10532},
  year={2018},
  publisher={Royal Society of Chemistry}
}

@article{walther2008janus,
  title={Janus particles},
  author={Walther, Andreas and M{\"u}ller, Axel HE},
  journal={Soft matter},
  volume={4},
  number={4},
  pages={663--668},
  year={2008},
  publisher={Royal Society of Chemistry}
}

@article{khaldy2020straight,
  title={Straight-line orientation in the woodland-living beetle Sisyphus fasciculatus},
  author={Khaldy, Lana and Tocco, Claudia and Byrne, Marcus and Baird, Emily and Dacke, Marie},
  journal={Journal of Comparative Physiology A},
  volume={206},
  pages={327--335},
  year={2020},
  publisher={Springer}
}
\end{document}